\documentclass[12pt,twoside]{article}

\usepackage{fleqn,espcrc1}

\newcommand{\AmS}{{\protect\the\textfont2
  A\kern-.1667em\lower.5ex\hbox{M}\kern-.125emS}}

\title{Quark degrees of freedom in hadronic systems}

\author{Vicente Vento \address{Departamento de F\'{\i}sica Te\'orica - IFIC,
        Universidad de Valencia - CSIC, \\ 
        E-46100 Burjassot (Valencia), Spain}%
        \thanks{The work reported here was supported by Spain 
                DGCYT-PB97-1127 and the European Union ERB 
                FMRX-CT96-008.}
}

\begin{document}

\maketitle

\begin{abstract}
Quantum Chromodynamics ($QCD$) is the theory of the strong interactions.
We review descriptions of hadronic systems motivated by $QCD$, analyzing
the recent controversy between gluonic and bosonic degrees of freedom
under the prism of the Cheshire Cat Principle. Our analysis leads to an
optimal scheme to study hadronic properties. We proceed to extend this
low energy descriptions to the deep inelastic regime.
\end{abstract}

\section{Introducing Quantum Chromodynamics}

There is little doubt nowadays in the physics community that
Quantum Chromodynamics ($QCD$) \cite{fgl,fy} is the theory that
describes the hadronic interactions within the Standard Model.
$QCD$ is a renormalizable quantum field theory whose elementary
fields are quarks, spin $\frac{1}{2}$ fermions with color ($A=1,2,3$)
and flavor ($i=1,\ldots,6$), and gluons, spin $1$ bosons with color
($a=1,\ldots,8$), coupled to achieve $SU_{color}(3)$ gauge
invariance of the theory,

\begin{equation}
{\cal L} = i\bar{\Psi}_i^A(x)\gamma^\mu (D_\mu)_{AB} \Psi(x)_i^B -
M_{ij}\bar{\Psi}_i(x)\Psi_j(x) -
\frac{1}{4} F^{\mu \nu a}F^a_{\mu \nu} +
\mbox{gauge fixing} + \mbox{ghosts}.
\label{qcd}
\end{equation}
The quark fields have been denoted by $\Psi_j^A$, the covariant derivative
is given by
\begin{equation}
(D_{\mu})_{AB}= \delta_{AB}\partial_{\mu} -igA_{\mu}^a (t^a)_{AB}
\end{equation}
where $t^a$ are the group generators in the quark representation satisfying
the algebra of the group, $[t_a,t_b] =f_{abc}t_c$, $A^a_{\mu}$ denote the
gluon fields and $g$ the coupling constant. The dynamics provides
self-couplings for the gluons as can be seen from the definition of
the gluonic tensor,
\begin{equation}
F_{\mu \nu}^a = \partial_{\mu}A^a_{\nu} - \partial_{\nu}A^a_{\mu} +
gf_{abc}A_{\mu}^bA_{\nu}^c.
\end{equation}
Finally the only dynamics associated with flavor comes from the mass
matrix,
\begin{equation}
M_{ij} = m_i\delta_{ij},
\end{equation}
a parameter set whose origin is external to the theory.

Despite its apparent simplicity, almost thirty years have passed since
its original formulation, and we have not yet been able to solve it exactly.
However much has been learned about its structure and its properties.

\subsection{Exact Results}
There has been progress in understanding the properties of strongly
interacting gauge theories through the development of rigorous
inequalities. In particular a surprising amount of information
about symmetry realization in $QCD$ follows from relatively simple
facts about fermion determinants and propagators in Euclidean space-time 
\cite{wvw} and anomaly constraints \cite{hooft}:

\begin{itemize}

\item [i) ] the pion mass is the smallest among the mesons;

\item [ii)] the pion mass is smaller than any baryon mass;

\item [iii)] if axial currents are conserved the axial symmetry
is spontaneously broken;

\item [iv)] if vector currents are conserved the vector symmetry is
unbroken;

\item [v)] parity is not spontaneously broken \cite{azcoiti}.

\end{itemize}

These exact results confirm that $QCD$ verifies the experimental observation
that chiral symmetry is realized in the hadronic interactions \`a la
Goldstone, i.e. in spontaneously broken fashion, and that the pion is the
lightest hadron in nature. Thus, even in the massless quark limit, the $QCD$
vacuum breaks chiral symmetry despite the fact that the currents derived
from the $QCD$ lagrangian are conserved.

\subsection{Asymptotic freedom}

A very important property of $QCD$, known as asymptotic freedom, is that
the effective coupling constant vanishes at short distances \cite{gwp}.
This can be mathematically expressed, in perturbation theory to leading
order, by the so called running coupling constant
\begin{equation}
\alpha_S (Q^2) = \frac{12\pi}{(33-2n_f) \log{(\frac{Q^2}{\Lambda^2})}}.
\label{alpha}
\end{equation}
A mathematical statement that implies that at large spacelike momenta
($Q^2 \rightarrow \infty$) the theory behaves as a free field theory
modulo logarithmic corrections. $\Lambda$ introduces the scale for
perturbative phenomena in the theory. Moreover for sufficiently large
momenta the corrections to the free field theory can be calculated in
a power series in $\alpha_S (Q^2)$. The contributions to the physical
processes are determined by the elementary vertices determined from the
lagrangian and the renormalization group.

These features can be seen directly by looking at high energy Jets 
\cite{jets} and have been proven extremely successful in the interpretation of
the Deep Inelastic lepton-hadron Scattering (DIS) data, what is known
as Bjorken scaling and the logarithmic deviations from it \cite{dis}.

\subsection{Confinement}

How does the theory behave away from the asymptotic  regime? We have not
been able to isolate quarks and gluons in the laboratory. This could have
been taken as a failure of $QCD$, however it has turned into the dynamical
principle which governs the low energy behavior of the theory known as
confinement. The fact that $\alpha(Q^2)$ grows for low momenta (see Eq.
(\ref{alpha})) is taken as an indication that quark and gluons are confined.
However the theory becomes in that regime highly non-perturbative and one
cannot use perturbative technology to describe this dynamics. Does
confinement arise from the quantum theory obtained from Eq(\ref{qcd})?
We have not been able to answer this question exactly except in particular
situations: lower dimensionality \cite{hooft1,albert}, Supersymmetric
Yang Mills theories \cite{misha} and in the most promising approximation
to the theory, i.e. Lattice QCD, for pure glue \cite{lattice}.

The problem with real QCD is that the structure of the vacuum is highly
non trivial. It is precisely the vacuum properties which dominate the low
energy behavior. All types of non perturbative field configurations inhabit
the groundstate, instantons, merons, monopoles,..., and moreover seem to
play a role in the way the theory is realized at the hadronic level.

Let us present a plausible confinement scenario \cite{mandelstam}.The
effective color interaction becomes stronger as the separation between
the probe color charges increases. In this regime a new phenomenon takes
place. When the distance becomes larger than some number times
$\frac{1}{\Lambda}$, and exceeds a critical one, the branching of the
gluons becomes so intensive that one cannot speak about individual gluons,
but rather should describe the interaction in terms of chromoelectric and
chromomagnetic fields. In $QCD$, the chromoelectric field between the probe
charges squeezes into a sausage-like configuration, the flux tube. This
situation is reminiscent of the Meissner effect in superconductivity. We
deal here with chromoelectric flux-tubes, i.e. with a Dual Meissner effect.
Confinement is insured by a condensation of magnetic monopoles (not charges),
which force the fields to form the flux tubes in the $QCD$ vacuum.

Nobody has ever proved this scenario. There are however indications arising
from lattice-$QCD$ calculations \cite{dual} and from N=2 SUSY Yang Mills
\cite{susy} where in the strong coupling regime monopoles have been shown
to condense.

\subsection{Chiral symmetry}
In the massless quark limit $QCD$ is invariant under $SU(n_f)\otimes
SU(n_f)$ chiral symmetry. This symmetry is however spontaneously broken
and for two flavors the pion is the associated Goldstone boson.
Microscopically the properties of the vacuum are crucial in describing
the realization of chiral symmetry. The Casher-Banks formula\cite{cb},
\begin{equation}
<\bar{q}q> = -\pi \rho(\lambda=0)
\end{equation}
tells us that the condensate, namely the order parameter for the transition
from the unbroken to the spontaneously broken phase, is connected with the
density ($\rho$) of quark states of small virtuality ($\lambda$). It has
been shown \cite{shuryak} that the instantons play an instrumental role
in explaining microscopically the realization of chiral symmetry.

\subsection{Imitating $QCD$}

The main effort in understanding $QCD$ at present, since an exact solution
seems non atainable, is into developing its lattice approximation. However
this approach is still very primitive in order to produce, without guidance,
realistic results. It may serve to support or reject some of the scenarios
that may be proposed in view of the data, but as of yet is still far away
from being able to predict or explain many of the hadronic properties
\cite{lattice}.

Physicists have turned to other means of understanding the data by
approximating the theory. The effective theory approach has been
widely used for describing low energy data. In this method one starts
from the original action and measure of the theory, integrates out
the irrelevant degrees of freedom and using the appropriate
matching conditions obtains a different, much more tractable, field
theory with the same S-matrix \cite{georgi}. This procedure has been used
with notable precision at high energies. At low energies we are not able to
follow strictly the procedure and therefore apply a celebrated, though unproven,
theorem by Weinberg \cite{weinberg}, which states that one should
write the most general lagrangian, constructed from the accesible
degrees of freedom in the domain under consideration, which satisfies
the relevant symmetries of the theory. This is
the scheme used for Chiral Perturbation Theory \cite{gl}, which
has been extremely successful in low energy meson physics.

A second procedure has been the use of approximation methods to truncate
$QCD$. The most relevant examples are $\frac{1}{N}$\cite{hooft2}
and $HQET$\cite{iw}. In this case we proceed in terms of an
expansion parameter $\frac{1}{N}$, $N$ being the number of colors for
the former, and $\frac{1}{M}$, $M$ being the mass of the heavy quark,
for the latter. Unluckily we have been only able to develop arguments in the
$\frac{1}{N}$ case since, despite many efforts, we have not been able to
sum the planar graphs for $QCD$. $HQET$ has
been extremely successful in discovering new spin symmetries for the heavy
quark sector and describing decays in terms of only a few parameters.

Lastly, people have resorted to use their imagination to produce models,
whose relation with the theory is at most qualitative, but which are extremely
predictive once they are fully developed. The basic assumption is that one
knows the way the theory is realized at low energies, i.e., the vacuum,
the Fock states and the dynamics among the constituents. There are many
models, which are becoming more detailed following the  advances in
our  microscopic understanding of the properties of the theory.

The subject of my discussion will be the physics associated with models
of hadron structure, but before I proceed along these lines let me 
mention an example of how specific features of the microscopic structure 
of the theory
can motivate explanations for particular phenomena.
The instantons building the
$QCD$ vacuum induce interactions among the quarks which have been shown
to be responsible for many observable effects of hadron physics,
both in the vacuum and in the nuclear
medium \cite{shuryak,kochelev,yasuo}.

It has become clear by now that it is hard to make definitive statements about
hadrons made up of light quarks (u,d and s), were the nonperturbative structure
of the theory is difficult to avoid. Nevertheless the hadron  spectrum and the
interactions among the hadrons display regularities which,  when properly
formulated, correlate with simple properties of the fundamental theory. They
are best expressed by inventing mechanisms that describe the undisclosed
features of the theory.

\section{Modelling hadron structure}

The role of models in $QCD$ is to produce simple physical pictures that
connect the phenomenological regularities with the underlying structure.
Models should be used to guide fundamental calculations and experiments.
They are not substitute for the theory and therefore should be closely
tied to it and abandoned when they become a divertimento by themselves.

Historically the first models of hadron structure appeared in the sixties
right after Gell-Mann and Zweig \cite{gmz} introduced the Quarks. Dalitz
in his beautiful Les Houches Lectures of 1965 \cite{dalitz}, performed a
phenomenological analysis leading to flavor forces of the type

\begin{eqnarray}
U(\bar{q} \; q) &= &U_{p1}(1 - \vec{\sigma_1}\cdot 
\vec{\sigma_2}) (1 - \tilde{F}_1\cdot\tilde{F}_2) 
+ U_{p8}(1 - \vec{\sigma_1}\cdot \vec{\sigma_2}) 
(8 + \tilde{F}_1\cdot\tilde{F}_2) + \nonumber \\
& &  U_{v1}(3 + \vec{\sigma_1}\cdot \vec{\sigma_2}) 
(1 - \tilde{F}_1\cdot\tilde{F}_2)  + U_{v8}(3 + \vec{\sigma_1}\cdot \vec{\sigma_2}) 
(8 + \tilde{F}_1\cdot\tilde{F}_2) 
\end{eqnarray} 
where $\vec{\sigma}$ arises from the quark spin generators and $\tilde{F}$ from
the SU(3) flavor generators. This most general Ansatz produced a beautiful fit,
which had to be dropped because no explanation was found to justify the
abandonment of the Spin-Statistics theorem. The introduction of color
\cite{ghn} solved all these problems and led to the formulation of $QCD$.

The Quark Model was revisited soon after the introduction of
$QCD$ by De R\'ujula, Georgi and Glashow \cite{rgg}, who developed a
scheme to incorporate color to the spectroscopic analysis. Their philosophy
has motivated many of the developments thereafter. In particular the
Bag Model \cite{mit} relies on a quantum field theoretic description
to implement the same philosophy. The aim of these approaches is to
define a scheme which incorporates the properties of $QCD$ as we next
describe.

\subsection{Confinement}

The models contain a mechanism which provides the dynamics to bind
quarks (antiquarks) to form color singlet hadrons. These mechanisms are
varied: color independent attractive potentials \cite{rgg,ik},
field boundary conditions providing the confinement of color to a
space-time tube \cite{mit}, etc.

\subsection{Asymptotic freedom}

Guided by the apparent absence of strong renormalization (higher twist
effects) in DIS it was thought that once confinement is implemented
a picture of a hadron arises where perturbative $QCD$ is qualitatively
reliable. Therefore an additional residual interaction between the quarks
based on the One Gluon Exchange contribution was introduced.
This so called color-magnetic interaction has an operator
structure in color space of the form

\begin{equation}
\sum_{i\ne j} F(r_{ij}) \lambda_i \cdot \lambda_j
 \vec{\sigma_i}\cdot \vec{\sigma_j}
\end{equation}
which has been implemented in all approaches and is responsible
among other things for the $\Delta$-Nucleon mass splitting. With
the proper simplifications, taking into account the Fock Space
of the lowest quark states, it can be reduced to an interaction
of flavor type \cite{jaffe} which reflects the strong flavor 
dependence explicit in the original parametrization of Dalitz.

Models satisfying these two properties have been extremely successful in
explaining a large amount of data \cite{ik,jaffe}. They have been
used as low energy descriptions of $QCD$, and by means of the Operator
Product Expansion, we have been able to study and predict high energy
data \cite{traini,scopetta}. The so called proton spin problem can be
understood as a transition from constituent quarks to currents quarks,
antiquarks and gluons. Moreover it has been shown that the data serve
to unveil the Regge structure of the constituent quarks \cite{scopetta}.
Moreover they have also been used to speculate about the behavior of
hadrons in the nuclear medium \cite{nimai}.

Despite the many successes of the schemes just described, they all lack
an ingredient, namely the realization of a fundamental symmetry
of $QCD$, i.e. spontaneously broken chiral symmetry. The role of the
pseudoscalar mesons as Goldstone boson was not incorporated in these models
and the pion is considered a conventional $\bar{q} q$ boundstate. This absence
implied that all traditional low energy hadron physics, dominated by chiral
flavor symmetry, had to appear from confined perturbative $QCD$. Although
the static properties were well reproduced, except in the pseudoscalar
meson sector, whose mass should vanish in the massless quark limit, the
attempts  to reproduce the long distance properties
of the hadronic interaction failed.

\subsection{Chiral symmetry}

Chiral symmetry was implemented initially in the bag model formalism
since the field theoretic language was more suitable for the task. Confined
perturbative $QCD$ was kept in the interior of the bag, while in the outside
region an elementary meson field was introduced by means of a non linear sigma
model. The dynamics between the quarks and the pions arises naturally from
implementing chiral current conservation at the boundary \cite{cbm}.
Soon thereafter attempts were made to implement this scheme
into the potential models \cite{vento}.

These models are able to reproduce the static properties also with
great precision. Moreover the appearence of the Goldstone modes allows
them to describe without further ingredients low energy properties associated
with chiral symmetry in a straightforward fashion. Let us describe two
lines of thought much developed in these days.

\subsubsection{The Cheshire Cat Principle}

In two dimensions fermion theories are exactly bosonizable. Let us
separate space arbitrarily in two regions. The left hand side we
describe in terms of a fermion theory, e.g. $QCD$ and the right hand
side in terms of the equivalent bosonic theory. If the appropriate
boundary conditions are defined, by matching both classical current
conservation and quantum mechanical anomalies, the Cheshire Cat
Principle (CCP) states: the physical properties have to be independent
of the position of the boundary \cite{nielsen}. In four dimensions
the situation can be best appreciated in a bag picture. Inside the bag we
have confined perturbative $QCD$ and outside the equivalent bosonic
theory, where solitons, i.e., skyrmions, carry the baryon number.
The CCP states that the physical observables have to be radius
independent \cite{rho}. $QCD$ in four dimensions is not exactly
bosonizable nor it is solvable to any degree of approximation in the
quark gluon language, thus the CCP principle can be only approximate.
We have used the opposite logic. We have taken the CCP as a quality
control check over the approximations involved \cite{rho}: only for those
values of the radius where there appears an almost radius independence
can the approximations be trusted. Monotonic functions 
of the radius for the observables imply that important contributions
are missing. Many observables have been analyzed successfully under this
philosophy. It is important to stress that not all the observables
require the same approximations. Those observables, where anomalies
contribute, like in the spin problem, are very delicate to calculate,
because no contribution is dominant and large cancellations among all
of them occur \cite{park}.

\subsubsection{Chiral quarks}

Manohar and Georgi tried to understand the successes of the non-relativistic
quark model in terms of effective theories \cite{mg}. Their main argument is
that the scale associated with confinement $\Lambda_{QCD}$
($\approx 100 - 300$ MeV) is smaller than that associated with
chiral symmetry breaking $\Lambda_{\chi SB}$ ($\approx 1000$ MeV).
Therefore there is a region of momentum where quarks, gluons and pions
coexist. Using the effective theory approach they obtain a theory where
quarks and gluons interact by means of the conventional color couplings
while quarks and pions through a non-linear sigma model. The quark and gluon
fields are effective fields, i.e., the latter with a constituent mass
obtained from chiral symmetry breaking. By naive dimensional arguments
and by properly matching the unknown coupling constants of the theory they
get from the hyperfine splitting $\alpha_s \approx 0.3$ and
from $\pi-\pi$ scattering the chiral scale $\Lambda_{\chi SB} = 4\pi f_\pi$, 
where $f_\pi \approx 93$ MeV is the pion decay constant.
They also show that gluon loops are suppresed as compared to quark and pion
loops, a statement which is much dependent on their dimensional arguments
and matching scheme.

Let me conclude this section by stating that the implementation of chiral
symmetry both in bag model schemes and in effective theories leads to a
theory which contains quarks, gluons and Goldstone mesons as the
natural degrees of freedom to describe the low energy properties. I will not
support therefore any of the sides of the recent Isgur-Glozman\cite{i,g}
debate but conclude that both extremes are unnatural. As shown very clearly
in the investigations on the CCP a full quark gluonic theory would imply
the consideration of higher order contributions making the calculation
impractical, while a full mesonic theory would need ultimately the
inclusion of many mesons, not only making the calculation
impractical, but filled with parameters to be fitted. In my opinion certain
features of the structure of hadrons appear in an extremely simple manner
in terms of the chromomagnetic interaction \cite{jaffe}, while others, in
particular the long range tale of the nucleon-nucleon interaction are easily
described by the pion \cite{cbm}. A scheme which incorporates both features,
like the Chiral Bag Model or the Chiral Quark Model, is most appropriate
to develop a model due to its simplicity.

\section{Applications}

I proceed to motivate and review some of the developments surrounding my
own research. The investigation aims at testing models of $QCD$ in
different regimes and circumstances to show what they teach us about the
true theory and analyze if they are helpful in predicting new data.

\subsection{Approximate bosonization} 

The $CCP$ has been tested in many instances with notable success
\cite{nielsen}. It has been observed at the level of topological
quantities, i.e., baryon charge fractionation \cite{gjaffe,zahed} 
and approximately at the level of non topological observables, i.e.,
masses, magnetic moments, etc \ldots \cite{park,jackson}. The
explicit manifestation of the $CCP$ in the latter is through some
type of minimum sensitivity principle in terms of the bag radius.
Moreover the mean value about which observables are not sensitive to
the radius corresponds to the confinement scale
$R\sim\frac{1}{\Lambda_{QCD}}$.

There is one case, the flavor axial singlet charge ($FSAC$),
were its implementation had not been successful beyond doubt and therefore
it has merit our attention \cite{park}.  From the phenomenological
point of view the $FSAC$ is associated with the $\eta^\prime$ and
therefore with the anomaly \cite{hooft1}. This observable is relevant for
what has been referred to as the {\it proton spin problem}.  In the chiral
bag model the formulation is very elaborate.  Confinement induces through
quantum effects a color anomaly, which leads to a surface coupling of the
$\eta^\prime$ with the gluon field. The latter induces a gauge non
invariant Chern-Simons current, whose expectation value we need to
calculate. We have shown that
the presence of the surface term generated by the proper matching of the
color anomaly with the surface gluon-$\eta^\prime$ coupling allows us to
obtain for this observable a value close to the data over a wide range of
bag radii. The CCP was instrumental in obtaining this agreement
\cite{park}.

\subsection{Chiral quarks}

We have seen that in an effective theory approach the light chiral quarks
appear if $\Lambda_{\chi SB} >> \Lambda_{QCD}$. In the region between the
two scales we have quarks and gluons interacting through the $SU(3)$ color
interaction. Since $SU(3)\otimes SU(3)$ global chiral symmetry is
spontaneously broken, there must be also an octet of Goldstone bosons
which are introduced in the effective theory also as fundamental fields.
All other hadrons are obtained as $qqq$ or $\bar{q}q$ bound sates.

We shall use for effective lagrangian below the chiral scale a linear
realization, although non-linear realizations can be immediatly developed,
i.e.

\begin{eqnarray}
{\cal L} &= & i\bar{\Psi}_i^A(x) \gamma^\mu (D_\mu)_{AB} \Psi(x)_i^B -
M_{ij}\bar{\Psi}_i(x)\Psi_j(x) -
\frac{1}{4} F^{\mu \nu a}F^a_{\mu \nu} + \nonumber \\
& & \bar{\Psi}_i(x) (\sigma \delta_{ij} + i\gamma_5\vec{\pi} \cdot
\vec{\tau}_{ij})\Psi_j(x) +
\frac{1}{2}\partial_{\mu}\sigma\partial^{\mu}\sigma +
\frac{1}{2}\partial_{\mu}\vec{\pi}\cdot\partial^{\mu}\vec{\pi}.
\label{cqm}
\end{eqnarray}

This is the fundamental support for many of the chiral quark models. The
quarks here are effective quarks and therefore the masses appearing in the
lagrangian are constituent masses.

Our aim has been to approach phenomenology from this scheme and we have
developed a non-relativistic picture which incorporates besides the
confinement potential, the OGEP, the OPEP and the OSEP and with these
scheme we have described both the spectra and the $N-N$ and $N-\Delta$
scattering channels\cite {gonzalez,garcilaso}. This most naive approach
is very successful.

\subsection{Constituent quarks and partons}

The models we have been discussing in the previous sections have been
motivated by mostly static properties and therefore pretend to reflect 
the non-perturbative aspects of the theory. In what follows we would like
to analyze high energy data by using them. The much spoken about success
of perturbation theory is a non predictive statement. The renormalization
group relates scales and therefore the success is based on experimental
circumstances where either the non-perturbative aspects are cancelled
between several observations, or data are fed into the theory and the
evolution from one scale to another is analyzed. We would like to be
predictive with the caveat that we are substituting  the true theory by
models.

\subsubsection{Parton distributions from  quark models}

The basic idea in this approach arises from rephrasing the OPE which
states that,
\begin{equation}
F_i^n(Q^2) = M^n_{ij}F_j^n(Q_0^2),
\end{equation}
i.e., the moments of  structure functions at one scale are related
by means of perturbatively calculable transformation matrices to the same
moments at another scale \cite{fy}. If $Q_0^2$ is taken to be a low scale,
what we have labelled hadronic scale, the F functions become highly non
perturbative matrix elements in general. We substitute these matrix
elements at the hadronic scale by the same matrix elements calculated with
a model. In particular we are able to relate the valence quark
distribution functions with the momentum
distributions in the corresponding baryonic state $n_q^a$, i.e. with the
hadronic wave functions of the model,
\begin{equation}
xq_V^a(x) = \frac{1}{(1-x)^2}\int d^3p \;n_q^a(\vec{p})\; \delta
(\frac{x}{1-x} - \frac{k_+}{M})
\end{equation}
where $a$ represents the diverse degrees of freedom (unpolarized,
$\uparrow$, $\downarrow$, $\ldots$), $p_+ = p_0 - p_z$, $x$ is the
Bjorken variable and $M$ the mass of the baryonic state.

In this way we have studied polarized and unpolarized structure function,
transversity distributions and angular momentum distributions with various
models \cite{traini,scopetta}. The results of our calculations show that
these models, with the parameters fixed by low energy properties are able
to provide a qualitative description of the data and therefore become
predictive. They are however too naive and new ingredients not seen by the
low energy probes have to be incorporated.

\subsubsection{Constituent and current quarks}

Our basic assumption above has been that gluon and sea bremsstrahlung are the
source of difference between the constituents and the current quarks. We
have gone beyond that scheme by incorporating some structure to the
quarks following the procedure we have called ACMP \cite{acmp}. Within
this approach constituent quarks are effective particles made up of
point-like partons (current quarks, antiquarks and gluons), interacting
by a residual interaction described by a quark model \cite{scopetta1}.The
structure of the hadron is obtained by a convolution of the constituent 
quark model wave function with the constituent quark structure function.
For a proton made up of $u$ and $d$ quarks,
\begin{equation}
f(x,\mu_0^2) = \int^1_x
\frac{dz}{z}[u_0(z,\mu_0^2)\Phi_{uf}(\frac{x}{z},\mu^0) +
d_0(z,\mu_0^2)\Phi_{df}(\frac{x}{z},\mu^0)],
\end{equation}
where $\mu_0^2$ is the hadronic scale, $f = q_v, q_s, g$ (valence
quarks, sea quarks and gluons respectively) and
$\Phi$ represents the constituents probability in each quark and has been
parametrized following general arguments of $QCD$ as
\begin{equation}
\Phi_{qf}(x,\mu_0^2) = C_f x^{a_f}(1-x)^{A_f-1}.
\end{equation}
The constants have been fixed by Regge phenomenology and the choice of the
hadronic scale ($\mu_0 = 0.34$ GeV$^2$). The discussion can be generalized
to the polarized structure functions. The procedure is able to reproduce
the data extremely well and in this framework the so called spin problem
does not arise.

\section{Concluding remarks}

$QCD$ has a complex non-perturbative realization which has escaped
solution for the past 27 years. The recent progress in calculating
directly from the theory has come from the lattice approximation, however
these calculations are still primitive, and certainly expensive, if one
wants to compare with data. In the meantime effective theories and models
should provide lattice $QCD$ with well defined scenarios where the theory
can be tested.

We have emphasized the relation between models and the properties of the 
theory. The concept of effective theory has been most instrumental for our
developments. Effective theories provide the parameters that one has to 
calculate from first principles. Models provide the scenarios where new physics
may be envisaged.

We have approached the recent controversy, about the most adequate degrees 
of freedom to describe hadron properties, from the point of view of the $CCP$.
Our conclusion is that there is a price to pay, namely complexity and lack of
predictivity, if one doest not consider gluon degrees of freedom to address
scenarios where asymptotic freedom is fundamental in determining the physics.
Moreover, there is no way, besides solving $QCD$, to describe without
pions scenarios where spontaneous broken chiral symmetry dominates the
dynamics. Our $CCP$ analysis has taught us that models, called generically
hybrid, are the most economical way to describe the physics of medium
energy.

We have shown that many models, developed to describe low energy
physics, if taken as approximate solutions of $QCD$ in the hadronic regime, are
able to describe, via evolution, properties of the asymptotic regime surprisingly
well and moreover be predictive.

\section*{Acknowledgements}

I have recalled some of the recent developments from the perspective of my own
personal research. Many friends, luckily too many to cite them here, have
accompanied me in the voyage. You may find some of them in the reference list.
Others were simply too generous to sign my papers. Of all of them I would like
to single out a few, because they have strongly motivated my work. Pedro
G\'onzalez has developed chiral quarks to its present beauty. Mannque Rho has
mantained my interest in the $CCP$ and has forseen much of its power. Marco
Traini has kept my faith in our early discovery: the relation between low
energy models and high energy data. Sergio Scopetta has been the driving force
behind our recent work in the asymptotic regime, which has confirmed our
expectations.  Nimai brought me back to naive models and showed me their
beauty. I miss him!


\begin{thebibliography}{9}



\bibitem{fgl} H. Fritzsch, M. Gell-Mann and H. Leutwyler, Phys. Lett {\bf B47}
(1973) 365.

\bibitem{fy} F.J. Yndurain, {\it The theory of quark and gluon interactions}
(Springer Verlag, Heidelberg, Germany 1999).

\bibitem{wvw} D. Weingarten, Phys. Rev. Lett. {\bf 51} (1983) 1830; 
E. Witten, Phys. Rev. Lett. {\bf 51} (1983) 2351;
C. Vafa and E. Witten, Nucl. Phys. {\bf 234} (1984) 173. 
 
\bibitem{hooft} G. 't Hooft, {\it Recent developments in gauge theories} (Eds.
G. 't Hooft {\it et al}, Plenum Press, New York 1980).

\bibitem{azcoiti} V. Azcoiti and  A. Galante, Phys. Rev. Lett. {\bf 83} 
(1999) 1518.

\bibitem{gwp} D.J. Gross and F. Wilczek, Phys. Rev. Lett. {\bf 30} (1973) 1323,
H.D. Politzer, Phys. Rev. Lett. {\bf 13} (1973 1346. 

\bibitem{jets} R.K. Ellis, W.J. Stirling and B.R. Webber, {\it QCD and collider
physics} (Cambridge University Press, Cambridge, UK 1996). 

\bibitem{dis} R.G. Roberts, {\it The structure of the proton} (Cambridge
University Press, Cambridge, UK 1990).

\bibitem{hooft1} G. `t Hooft, Nucl. Phys. {\bf b138} (1978) 1.

\bibitem{albert} A. Ferrando and V. Vento, Phys. Rev. {\bf D49} (1994) 3044.

\bibitem{misha} M. Shifman, Int. J. Mod. Phys. {\bf A14} (1999) 5017. 

\bibitem{lattice} M. Creutz, {\it Quarks, gluons and lattices} (Cambridge
University Press, Cambridge, UK 1983).

\bibitem{mandelstam} S. Mandelstam, Phys. Rep. {\bf 23} (1976) 245; G. 't Hooft,
{\it Under the spell of the gauge principle} (World Scientific, Singapore 1994).

\bibitem{dual} A. Di Giacomo, Nucl. Phys. Proc. Suppl. {\bf 47} (1996) 136.

\bibitem{susy} N. Seiberg and E. Witten, Nucl. Phys. {\bf B426} (1994) 19, (E)
{\bf B439} (1994) 485; Nucl. Phys. {\bf B431} (1994) 484.

\bibitem{cb} T. Banks and A. Casher, Nucl. Phys. {\bf B168} (1980) 103.

\bibitem{shuryak} T. Sch\"afer and E. V. Shuryak, Rev. of Mod. Phys. {\bf 70}
(1998) 323.

\bibitem{georgi} H. Georgi, {\it Weak interactions and modern particle physics}
(The Benjamin/Cummings Pub. Co., Menlo Park, CA, 1984).

\bibitem{weinberg} S. Weinberg, Physica {\bf 96A} (1979) 327.

\bibitem{gl} J. Gasser and H. Leutwyler, Ann. Phys. {\bf 158} (1984) 142.

\bibitem{hooft2} G. `t Hooft, Nucl. Phys. {\bf B72} (1974) 461; E. Witten, Nucl.
Phys. {\bf B160} (1979) 57.

\bibitem{iw} N. Isgur and M. Wise, Phys. Rev. {\bf D43} (1991) 819.

\bibitem{kochelev}  N.I. Kochelev, V. Vento and A.V. Vinnikov,
Phys. Lett. {\bf B472} (2000) 247. 

\bibitem{yasuo} Y. Umino and V. Vento, Phys. Lett. {\bf B472} (2000) 5. 

\bibitem{gmz} M. Gell-Mann, Phys. Lett. {\bf 8} (1964) 214; G. Zweig, CERN
Preprints 401 and 402 (unpublished).

\bibitem{dalitz} R.H. Dalitz, {\it High energy physics} (Ed. C. de Witt and M.
Jacob, Gordon and Breach Science Pub., New York 1965).

\bibitem{ghn} O.W. Greenberg, Phys. Rev. Lett. {\bf 13} (1964) 598; M. Han and
Y. Nambu, Phys. Rev. {\bf 139} (1965) 1006.

\bibitem{rgg} A. de R\'ujula, H. Georgi and S.L. Glashow, Phys. Rev. {\bf D12}
(1975) 147.

\bibitem{mit} T. DeGrand, R.L. Jaffe, K. Johnson and J. Kiskis, Phys. Rev. {\bf
D12} (1975) 2060.

\bibitem{ik} N. Isgur and G. Karl, Phys. Rev. {\bf D20} (1979) 1191, {\bf D21}
(1979) 3175.

\bibitem{jaffe} R.L. Jaffe, Color, spin and color dependent forces in Quantum
Chromodynamics, hep-ph/001123.
 
\bibitem{traini}  M. Traini, L. Conci and U. Moschella, Nucl. Phys. {\bf A544}
(1992) 731; M. Ropele, M. Traini and V. Vento, Nucl. Phys. {\bf A584}
(1995) 634; M. Traini, A. Mair, A. Zambarda and V.
Vento, Nucl. Phys. {\bf A614} (1997) 472. 

\bibitem{scopetta}S. Scopetta and V. Vento, Phys. Lett. {\bf B424}(1998)25,
 Phys. Lett. {\bf B460} (1999) 8, Phys. Lett. {\bf B474} (2000) 235. 
 
\bibitem{nimai}  N.C. Mukhopadhyay and V. Vento Nucl. Phys. {\bf A643}(1998)
415. 

\bibitem{cbm} A. Chodos and C.B. Thorn, Phys. Rev. {\bf D12} (1975) 2733; 
V. Vento, M Rho, E. M. Nyman, J.H. Jun, G.E. Brown, 
Nucl. Phys. {\bf A345} (1980) 413;
G.E. Brown, A.D. Jackson, M. Rho, V. Vento, Phys. Lett. {\bf B140} (1984)
285; A.W. Thomas, S. Theberge, G.A. Miller. 
Phys. Rev. {\bf D24}(1981) 216. 

 
\bibitem{vento} J. Navarro and V. Vento, Phys. Lett. {\bf B140} (1984) 6,
 Nucl. Phys. {\bf A440} (1985) 617; Yu. A. Kuperin, A. A. Kvitsinsky, S. P.
Merkurev, E. A. Yarevsky; Sov. J. Nucl. Phys. {\bf 51}(1990) 141, 
Nucl. Phys. {\bf A523} (1991) 614. 

\bibitem{nielsen} S. Nadkarni, H.B. Nielsen and I. Zahed, Nucl. Phys. {\bf B253}
(1985) 308.

\bibitem{rho} B.-Y. Park, V. Vento, M. Rho and G. E. Brown, 
Nucl. Phys {\bf A504} (1989) 829.
 
\bibitem{park} H.-J. Lee, D.-P. Min, B.-Y. Park, M. Rho and V. Vento,
Nucl. Phys. {\bf  A657} (1999) 75. 

\bibitem{mg} A. Manohar and H. Georgi, Nucl. Phys. {\bf 234} (1984) 189.
 
\bibitem{i} N. Isgur, Critique of a pion exchange model for interquark forces,
 nucl-th/9908028. 

\bibitem{g} L.Ya. Glozman, Reply to Isgur's 'critique of a pion exchange model 
for interquark forces', nucl-th/9909021. 

\bibitem{gonzalez} A. Valcarce, P. Gonz\'alez, F. Fern\'andez and V. Vento,
Phys. Lett. {\bf B367} (1996) 35, Phys. Rev. {\bf C61} (2000) 019803; 
F. Fern\'andez, P. Gonz\'alez and A. Valcarce,
Few Body Systems Suppl. {\bf 10} (1999) 395; D.R. Entem, F. Fern\'andez and A.
Valcarce, Phys. Rev. {C62} (2000) (to be published).

\bibitem{garcilaso} H. Garcilazo, Abstracts of this conference and to be
published.
 

 
\bibitem{gjaffe} J. Goldstone and R.L. Jaffe, Phys. Rev. Lett. {\bf 51} (1983)
1518.


\bibitem{zahed} I. Zahed and G.E. Brown, Phys. Rep. {\bf 142} (1986) 1; M.
Nowak, M. Rho and I. Zahed, {\it Chiral nuclear dynamics} (World Scientific,
Singapore 1996).

\bibitem{jackson}  A.D. Jackson, L. Vepstas, E. W\"ust and D. Kahana, Nucl.
Phys. {\bf A462} (1987) 661.


\bibitem{acmp} G. Altarelli, N. Cabbibo, L. Maiani, R. Petronzio, Nucl. Phys.
{\bf B} (1974) 531.

\bibitem{scopetta1} S. Scopetta, V. Vento and M. Traini, Phys. Lett. {\bf 421}
(1998) 64; Phys. Lett. {\bf B442} (1998) 28.


\end{thebibliography}
\end{document}